\documentclass{tran-l}

\theoremstyle{definition}

\theoremstyle{remark}

\numberwithin{equation}{section}

\begin{document}

\title[Energy Distribution of a  G\"{o}del-Type Space-Time]
 {Energy Distribution of a G\"{o}del-Type Space-Time}

\author{ Ragab M. Gad}
\address{ Mathematics Department, Faculty of
 Science, Minia University, El-Minia, Egypt}

\email{ragab2gad@hotmail.com}






\dedicatory{}


\large{
\begin{abstract}
We calculate the energy and momentum distributions associated with
a G\"{o}del-type space-time, using the well-known  energy-momentum
complexes  of Landau-Lifshitz and M{\o}ller.  We show  that the
definitions of Landau-Lifshitz and M{\o}ller do not furnish a
consistent result.
\end{abstract}

\maketitle
\section{Introduction}
One of the most active areas of research in the general theory of
relativity is the energy-momentum complex. There are many attempts
to evaluate the energy distribution in a general relativistic
system beginning with the Einstein's energy-momentum complex and
followed by many prescriptions, for instance,  Landau-Lifshitz
(LL), Papapetrou and Weinberg (See in \cite{V99,ACV96} and
references therein). Most of these prescriptions only giving
meaningful results if the calculations are performed in
``Cartesian coordinates''. M{\o}ller \cite{M58} constructed an
expression which enables one to evaluate energy in any coordinate
system.
\par
Several examples of particular space-times (the Kerr-Newman, the
Einstein-Rosen and the Bonnor-Vaidya) have been investigated and
different energy momentum complexes are known to give the same
energy distribution for a given space-time \cite{1} - \cite{7}.
Vagenas \cite{Vagenas} has proved that in (2+1)d all complexes
give the same results. Virbhadra and Parikh \cite{8,9} studied the
energy distribution associated with a stringy charged black-hole
in Einstein's prescription. The same result has been obtained by
Xulu \cite{10} using Tolman's energy-momentum complex that is in
fact the Einstein energy-momentum complex   expressed by Tolman in
a different form (see \cite{1} for details.)

\par
Virbhadra \cite{V99}  noted that the definitions of LL, Papaperrou
and Weinberg give the same result as in the Einstein prescription
if the calculations are performed in Kerr-Schild Cartesian
coordinates. However, these complexes disagree with Einstein
definition if computations are done in  ``Schwarzschild  Cartesian
coordinates \footnote{ Schwarzschild metric in ``Schwarzschild
Cartesian coordinates'' is obtained by transforming this metric
(in usual Schwarzschild coordinates $\{t, r, \theta, \phi\}$)  to
$\{t,x,y,z\}$ using $ x = r \sin\theta \cos\phi, x = r \sin\theta
\sin\phi, z = r \cos\theta $.}.''

Recently, Xulu \cite{grO0} obtained the energy distribution for
the most general non-static spherically symmetric metric using
M{\o}ller's definition. He found different results in general from
those obtained using the Einstein energy momentum complex; these
results agree for the Schwarzschild, Vaidya and
Janis-Newman-Winicour space-times, but
disagree for the Reissner-Nordstr\"{o}m space-time.\\
Sharif \cite{Shraif} evaluated the energy and momentum density
components for G\"{o}del-type metric by using prescriptions of
Einstein and Papapetrou. He found that the two prescriptions
differ in general for this space-time. Dabrowski \cite{Dabrowski}
has calculated the energy  momentum and angular momentum, using
Einstein and Bergmann definitions, for both acausal and causal
G\"{o}del model.

\par
In this paper we will evaluate the energy and momentum densities
for a
 G\"{o}del-type space-time using the energy-momentum
complexes of Landau-Lifshitz and M{\o}ller. We obtain them
 if the space-time under consideration is
homogeneous. We are also interested to check whether or not the
Cooperstock hypothesis \cite{Coop} ( which states that the energy
and momentum in a curved space-time are confined to the regions of
non-vanishing energy-momentum tensor $T^{k}_{i}$ of the matter and
all non-gravitational fields) holds for this case.
\\ Through this paper we use $G = 1$ and $c = 1$ units
and follow the convention that Latin indices take value from 0 to
3 and Greek indices take value from 1 to 3.

\section{G\"{o}del-type space-time}
An exact solution of Einstein's field equations with cosmological
constant for incoherent matter with rotation has been found by
G\"{o}del in 1949 \cite{G49}. This solution, called G\"{o}del-type
manifold, described by the line element \cite{RT83}
\begin{equation}\label{GS}
ds^2 = \big[ dt + H(r)d\phi\big]^2 - D^2(r)d\phi^2 - dr^2 - dz^2,
\end{equation}
where $r, \phi$ and $z$ are cylindrical coordinates.\\
The non-zero components of the energy-momentum tensor for the
G\"{o}del-type metric (\ref{GS}) are
\begin{equation}\label{2.1}
\begin{split}
T_{0}^0 & = -\frac{1}{8\pi}\big(\frac{D^{\prime\prime}}{D}-
3\big(\frac{H^{\prime}}{2D}\big)^2
-\frac{H}{2D}\big(\frac{H^{\prime}}{D}\big)^{\prime}\big),\\
T_{2}^0 & = -\frac{1}{8\pi}\big[\frac{HD^{\prime\prime}}{D}
-\frac{D}{2}\big(\frac{H^\prime}{D}\big)^\prime -
\frac{3H}{4}\big(\frac{H^\prime}{D}\big)^2-
\frac{H^2}{2D}(\frac{H^\prime}{D})^\prime
- \frac{H^3}{D^2}(\frac{H^\prime}{2H})^2\big],\\
T_{1}^1 & = -\frac{1}{8\pi}\big(\frac{H^{\prime }}{2D}\big)^2,\\
T_{2}^2 & =
-\frac{1}{8\pi}\big[\frac{H}{2D}(\frac{H^\prime}{D})^\prime +
+\frac{H^2}{D^2}(\frac{H^\prime}{2H})^2\big],\\
T_{3}^3 & = -\frac{1}{8\pi}\big[-\big(\frac{H^\prime}{2D}\big)^2 +
\frac{D^{\prime\prime}}{D}\big].
\end{split}
\end{equation}
\par
The G\"{o}del cosmological solution has a well-recognized
importance and has to a large extent motivated the investigations
on rotating cosmological space-time within the frame work of
general relativity. Particularly, the search for rotating
G\"{o}del-type space-times has received a good deal attention in
recent years, and the literature on these geometries is fairly
large today (See Refs. \cite{SR68} - \cite{S95} and references
therein). \\ The homogeneity of this space-time was considered for
the first time in 1980 by Raychandhuri and Thakurta \cite{RT80}.
The necessary and sufficient conditions for a G\"{o}del-type
manifold (\ref{GS}) to be space-time homogeneous were proved in
\cite{TRA85}. These space-times are characterized by metrics of
the form (\ref{GS}) which, in addition, satisfy the following
restrictions
\begin{equation}\label{GC}
\frac{D^{\prime\prime}}{D} = m^2 = \text{const.} \quad \text{and}
\quad \frac{H^{\prime}}{D} = \text{const.}= 2\Omega,
\end{equation}
where prime denotes the differentiation w.r.t. $r$.
\par

 It is well known that the energy-momentum
complexes give meaningful result if they are evaluated in
Cartesian coordinates. According to the following transformations
$$
 x = r \cos\phi,
\quad y = r \sin\phi, \quad z = z, $$ and $$ r = \sqrt{x^2 + y^2},
$$ the line element (\ref{GS}) may be transformed to
quasi-Cartesian coordinates:

$$ ds^2 =dt^2 + 2H(r)\big(
\frac{x}{r^2}dy - \frac{y}{r^2}dx\big)dt - \frac{(D^{2}(r) -
H^{2}(r))}{r^4}(xdy - ydx)^2 $$

\begin{equation} \label{2.3}
- \frac{1}{r^2}(xdx + ydy)^2 - dz^2.
\end{equation}

The determinant of the metric tensor is
\begin{equation}\label{2.5}
g = -\frac{D^2}{r^2}.
\end{equation}
The components of the contravariant metric tensor are
\begin{equation}\label{2.6}
\begin{split}
g^{00} & = \frac{D^2 - H^2}{D^2}, \\ g^{01} & = -
\frac{yH}{D^2},\\ g^{02} & = \frac{xH}{D^2},\\ g^{11} & = -
\frac{x^2D^2 + y^2r^2}{r^2D^2},\\ g^{12} & = - \frac{xy(D^2 -
r^2)}{r^2D^2},\\ g^{22} & = - \frac{y^2D^2 + x^2r^2}{r^2D^2},\\
g^{33} & = -1.
\end{split}
\end{equation}
\section{Energy and momentum in Landau-Lifshitz's Prescription}
The energy and momentum densities in the sense of Landau-Lifshitz
\cite{LL} are given by

\begin{equation}\label{P0}
P^{m} = \frac{1}{16\pi}S^{mj0k}_{,jk},
\end{equation}
where
\begin{equation}\label{}
S^{mjnk} = -g\big( g^{mn}g^{jk} - g^{mk}g^{jn}\big),
\end{equation}
and has symmetries of the Riemann tensor.\\
$P^0$ is the energy density and $P^\alpha$ are the momentum
density components.
\par
In order to calculate the energy and momentum densities for
G\"{o}del-type expressed by the line element (\ref{GS}) we need
the following non-zero components of $S^{mjnk}$
\begin{equation}\label{S}
\begin{split}
S^{0101} & = \frac{x^2}{r^4}(H^2 - D^2) - \frac{y^2}{r^2}, \\
S^{0102} & = \frac{xy}{r^4}(H^2 - D^2) + \frac{xy}{r^2},\\
S^{0202} & = \frac{y^2}{r^4}(H^2 - D^2) - \frac{x^2}{r^2},\\
S^{1201} & =\frac{xH}{r^2},\\
S^{1202} & =  \frac{yH}{r^2}.
\end{split}
\end{equation}
Substituting (\ref{S}) in (\ref{P0}), one gets the energy and
momentum densities in Landau-Lifshitz as
$$
P^{0}  = \frac{1}{8\pi r^4}\big( (H^2 -D^2) - 3r(HH^\prime -
DD^\prime) + r^2(H^{\prime 2} - D^{\prime 2} + HH^{\prime\prime} -
DD^{\prime\prime})\big),
$$
\begin{equation}\label{2.6}
\begin{split}
P^{x} & = \frac{y}{16\pi r^3}(rH^{\prime\prime} - H^\prime),\\
P^{y} & = \frac{x}{16\pi r^3}( H^\prime - rH^{\prime\prime} ),\\
P^{z} & = 0.
\end{split}
\end{equation}
\par
The line element (\ref{GS}) with  $H = e^{ar}$, $D =
\frac{e^{ar}}{\sqrt{2}}$, $a = - m$, and $m^2 = 2\Omega^2$ is
homogenous in space and time (hereafter called ST-homogeneous).
Hence, from equation (\ref{2.1}), the non-vanishing components of
$T^a_b$ become.
\begin{equation}\label{2.11}
\begin{split}
T_{0}^0 & = \frac{\Omega^2}{8\pi},\\
T_{2}^0 & = \frac{\Omega^2}{4\pi},\\
T_{1}^1 & = -\frac{\Omega^2}{8\pi},\\
T_{2}^2 & =-\frac{\Omega^2}{8\pi},\\
T_{3}^3 & = -\frac{\Omega^2}{8\pi}.
\end{split}
\end{equation}
 From equations (\ref{2.6}) the energy and momentum
densities for ST-homogeneous become
\begin{equation}\label{2.7}
\begin{split}
P^0 & = \frac{e^{2ar}}{16\pi r^4}(r-1)(2r-1),\\ P^x &
=\frac{aye^{ar}}{16\pi r^3}(ar-1),\\ P^y & = \frac{axe^{ar}}{16\pi
r^3}(1-ar).
\end{split}
\end{equation}
It is worthy to investigate the Cooperstock hypothesis for the
ST-homogeneous. We see that when $r = 0$, the components of energy
momentum tensor (\ref{2.11}) have finite values, while from the
equations (\ref{2.7}) the energy and momentum components tend to
infinity. Therefore the results obviously not uphold the
Cooperstock hypothesis.

\section{Energy Distribution in M{\o}ller's Prescription}
M{\o}ller's energy-momentum complex \cite{M58} is given by
\begin{equation}\label{4.1}
\Theta^k_i = \frac{1}{8\pi}\chi ^{kl}_{i,l},
\end{equation}
where the antisymmetric superpotential $\chi^{kl}_i$ is
\begin{equation}\label{4.2}
\chi^{kl}_i = - \chi^{lk}_i = \sqrt{-g}\big( \frac{\partial
g_{in}}{\partial x^m} - \frac{\partial g_{im}}{\partial x^n}\big)
g^{km}g^{nl},
\end{equation}
$\Theta^0_0$ is the energy density and $\Theta^0_{\alpha}$ are the
momentum density components. \\
Also, the energy-momentum complex $\Theta^{k}_{i}$ satisfies the
local conservation laws:
\begin{equation}\label{4.3}
\frac{\partial\Theta^k_i}{\partial x^k} = 0
\end{equation}

The only non-vanishing components of $\chi^{kl}_{i}$ are
\begin{equation}\label{4.4}
\begin{split}
\chi^{01}_{0} & = \frac{HH^\prime}{D},\\
\chi^{01}_{2} & = \frac{1}{D}\big(H^\prime(H^2 + D^2)-2DD^\prime
H\big).
\end{split}
\end{equation}
Using the above components in (\ref{4.1}), we obtain the energy
and momentum densities in the form
\begin{equation}\label{4.5}
\begin{split}
\Theta^0_0 & = \frac{1}{8\pi D^2}\big( DHH^{\prime\prime} + DH^{\prime 2} - HH^\prime D^\prime\big),\\
\Theta^0_2 & = \frac{1}{8\pi D^2}\big[ \big(H^2 +
D^2\big)\big(DH^{\prime\prime} - H^\prime D^\prime\big) +
2HD\big(H^{\prime 2} - DD^{\prime\prime}\big],\\
\Theta^0_1 & = \Theta^0_3 = 0.
\end{split}
\end{equation}
For the ST-homogeneous G\"{o}del-type space-time, the energy and
momentum densities become
\begin{equation}\label{4.6}
\begin{split}
\Theta^0_0 & = \frac{m^2 e^{-mr}}{4\sqrt{2}\pi},\\
\Theta^0_2 & = \frac{m^2 e^{-2mr}}{4\sqrt{2}\pi },\\
\Theta^0_1 & = \Theta^0_3 = 0.
\end{split}
\end{equation}
We note that from equation (\ref{4.6}), if $r$ tends to infinity
the energy and momentum components tend to zero, but from equation
(\ref{2.11}) the energy momentum tensor componebts, $T^a_b$, have
finite values. Therefore the M{\o}ller's definition not upholds
the Cooperstock hypothesis for the ST-homogeneous.

\section{Discussion}

Ever since the first investigations of cylindrically symmetric
space-times by Levi-Civita \cite{L-C17} and , latter by Lewis
\cite{L32}, these space-times have been studied extensively for
their mathematical and physical properties. These have recently
been studied particularly in context of black hole \cite{L9598},
gravitational waves and cosmic strings \cite{H0}. Some examples of
well known cylindrically symmetric astrophysical and cosmological
solutions discussed in the literature include Einstein-Maxwell
fields \cite{L95}, magnetic strings \cite{DL02}, static
gravitational fields \cite{CNV97} and a large number of cosmic
string solutions.
\par
Using different definitions of energy-momentum complex, several
authors  studied the energy distribution for a given space-time.
Most of them restricted their intention to the static and
non-static spherically symmetric space-times.

\par
 In the present
work we investigated the energy and momentum associated with the
space-time in cylindrical coordinates using Landau-Lifshitz and
M{\o}ller definitions. We obtained the energy and momentum density
components of the  G\"{o}del-type space-time as well as the
homogeneous G\"{o}del-type space-time in these two prescriptions.
We found that the two definitions of energy-momentum
complexes do not provide the same result for these type of metrics.\\
The homogenous G\"{o}del-type space-time provide one more example
where the energy and momentum localization together with the
Cooperstock hypothesis are far from resolved.

}
\end{document}